\documentstyle[prl,aps,preprint,epsfig]{revtex}

\newlength{\figwidth}
\multiply\figwidth by 4
\divide\figwidth by 10

\begin{document}

\preprint{Radiation Trapping}

\title{Radiation trapping in coherent media}

\author{
 A.~B.~Matsko,
 I.~Novikova,
 M.~O.~Scully,
 and G.~R.~Welch
}
\address{%
 Department of Physics,
 Texas A\&M University,
 College Station, Texas 77843-4242,%
}

\date{\today}

\maketitle

\begin{abstract}

	We show that the effective decay rate of Zeeman
coherence, generated in a $^{87}$Rb vapor by linearly polarized
laser light, increases significantly with the atomic density.
We explain this phenomenon as the result of radiation trapping.
Our study shows that radiation trapping must be taken into
account to fully understand many electromagnetically induced
transparency experiments with optically thick media.

\end{abstract}


\newpage

\narrowtext

	Coherent population trapping (CPT) was first observed
in experiments establishing Zeeman coherence in sodium
atoms~\cite{first}.  In these experiments, explained in terms
of a three-level $\Lambda$-type level scheme, a laser field was
used to create superpositions of the ground state sublevels.
One of these superpositions, referred to as the ``bright''
state, can interact with the laser field while the other
superposition does not and is referred to as the ``dark''
state~\cite{Arimondo}.  All the population in the system is
eventually optically pumped into the dark state, and resonant
absorption of the electromagnetic field almost disappears.
This phenomenon is one manifestation of electromagnetically
induced transparency (EIT)~\cite{Arimondo,EIT}.

	EIT is particularly interesting because it offers a wide
variety of applications ranging from lasers without population
inversion~\cite{LWI,phaseonium} to new trends in nonlinear
optics~\cite{new}.  One of the most striking phenomena connected
with EIT is that atoms prepared in a coherent superposition
of states can produce a steep dispersion and a large resonant
index of refraction with vanishing absorption~\cite{index}.
Preparation of matter in such a state (which has been dubbed
``phaseonium''~\cite{phaseonium}) provides us with a new type
of optical material of interest both in its own right, and in
many applications to fundamental and applied physics.

	A common condition for applications of EIT is a high
optical density of the resonant medium.  For example, in
experiments demonstrating enhancement of index of refraction
($\chi'\sim 10^{-4}$) the density of the particles was $N\sim
10^{12}~\mathrm{cm}^{-3}$~\cite{index}.  Also, for highly
sensitive magnetometry based on atomic phase coherence
the density of atoms is estimated to be $N \sim 5 \times
10^{12}~\mathrm{cm}^{-3}$~\cite{Scully_magn}.

	For optically thick media, reabsorption of
spontaneously emitted photons can become important.
This process, called radiation trapping, has been
studied extensively in astrophysics, plasma physics,
and atomic spectroscopy~\cite{Molish}.  Radiation
trapping has been predicted and demonstrated to have a
destructive effect on the orientation produced by optical
pumping~\cite{happer72rmp,rad_trap,rad_trap1}.  Because the
spontaneously emitted photons are dephased and depolarized
with respect to the coherent fields creating the atomic
polarization, the effect of radiation trapping can be
described as an { \em external incoherent pumping } of the
atomic transitions~\cite{rad_trap1}.  Under the conditions of
EIT, there are not many atoms undergoing spontaneous emission.
However, these spontaneous photons destroy the atomic coherence
in the same way as incoherent pumping.  This effect can change
the results of CPT and EIT experiments significantly.

	In this Letter we report the observation and
analysis of the increase of the effective decay rate of
Zeeman coherence due to radiation trapping.  We create the
coherence between ground state sublevels of the $\mathrm{D}_1$
line of ${}^{87}\mathrm{Rb}$ vapor, and study the effect of
CPT as a function of the vapor density using polarization
spectroscopy based on nonlinear magneto-optic rotation
(NMOR)~\cite{nl_Faraday,budker_sens,Fleischhauer}.  We see that
the the relaxation time of the coherent state is determined
not only by the time-of-flight of the atom through the laser
beam, but also by the density of the atomic vapor.  For atomic
densities $N \approx 5 \times 10^{12}~\mathrm{cm}^{-3}$
the effective coherence decay rate increases by several
times compared with the decay rate for $N \approx 5 \times
10^{11}~\mathrm{cm}^{-3}$.  This demonstrates the importance
of radiation trapping for experiments with optically thick
coherent media, and the need to account for it to understand
the experimental results.

	Our experimental setup is shown schematically in
Fig.~\ref{levels.fig}.  An external cavity diode laser is
tuned in the vicinity of the $F=2 \rightarrow F'=1$ transition
of the ${}^{87}\mathrm{Rb}$ $\mathrm{D}_1$ line.  The laser
beam passes through a high-quality polarizer $P_1$, and then
through a cylindrical glass cell containing isotopically
enhanced ${}^{87}\mathrm{Rb}$.  The length of the cell is
$L=5.0~\mathrm{cm}$ and its and diameter is $D=2.5~\mathrm{cm}$.
The laser power after the polarizer is $P=2.5~\mathrm{mW}$,
and the beam diameter is $d=2~\mathrm{mm}$.  The glass cell
is placed inside a two-layer magnetic shield to suppress
the laboratory magnetic field.  A homogeneous longitudinal
magnetic field is created by a solenoid mounted inside the inner
magnetic shield.  The density of the Rb vapor is controlled
with the temperature of the cell.  A second polarizer $P_2$
(a polarizing beam splitter) is placed after the cell and
is tilted at $45^\circ$ with respect to the first polarizer.
The two beams emerging from the polarizing beam splitter are
detected with detectors $D_1$ and $D_2$.  A simple analysis of
the signals from the two channels gives the angle of rotation
of the polarization $\phi$ and the transmitted intensity
$I_{\mathrm{out}}$.

	We consider the linearly polarized light as two
circular components $E_+$ and $E_-$ which generate a coherent
superposition of the Zeeman sublevels (a dark state).  To study
this dark state we apply a longitudinal magnetic field in the
direction of light propagation which leads to a splitting of
the $|b_\pm\rangle$ states of magnitude $\hbar\delta_0/2 =
- \mu_B B$ where $B$ is the magnetic field, $\mu_B$ is the
Bohr magneton.  Because the two circular components interact
with ground-state sublevels of oppositely signed magnetic
quantum number, the changes in the index of refraction for the
two components have opposite sign.  As a result of the index
change, the components acquire a relative phase shift $\phi$
which leads to a rotation of the polarization direction which
is proportional to the magnetic field for small fields.  Thus,
by measuring the rotation as a function of applied magnetic
field, we obtain information about the dispersive properties of
the medium.  As the magnetic field increases, the absorption
also increases because the splitting between the sublevels
destroys the two-photon resonance, so the detection of the
transmitted electromagnetic field intensity as a function of
the magnetic field allows us to study the absorptive properties
of the medium.  Hence, using the nonlinear Faraday technique
we can easily study both dispersive and absorptive properties
of the EIT resonance simultaneously.

	We have measured the polarization
rotation slope $d\phi/dB$ and the transmission
$I_{\mathrm{out}}/I_{\mathrm{in}}$ of the cell.  The detailed
shape depends on the particular cell and laser beam size,
but a typical result is shown in Fig.~\ref{fig2.fig}.
The individual points on this plot correspond to different
atomic densities.  These data cannot be fit by existing
theoretical considerations for this system assuming a constant
decay rate $\gamma_0$~\cite{Fleischhauer}.  This fact is
demonstrated by extrapolating the low density end of the curve
(where $I_{\mathrm{out}}/I_{\mathrm{in}}$ is nearly 1) with a
constant decay rate, as shown in the dashed curve and inset.
However, taking into account the effect of radiation trapping
we can understand this data quite well.

	We first analyze these data in terms of a simple
theoretical model and then by a detailed numerical simulation.
To simplify the analysis we neglect the process of optical
pumping and assume a closed system.  We include a dephasing
of the ground-state coherence with rate $\gamma_0$ and
neglect population exchange between the ground states.
As in previous treatments, we model the effect of radiation
trapping by introducing an incoherent pumping rate $R$ from
ground states $|b_\pm\rangle$ to excited state $|a\rangle$.
$R$ is not a constant, but a function of all parameters of
the system~\cite{rad_trap1}.  In particular, $R$ depends
on the excited-state population, because as more population
is transferred to the excited state, more radiation will be
produced that can eventually cause this incoherent excitation.
We assume that $R$ may be of the same order of magnitude
as $\gamma_0$ but that it is much less then the radiative
decay rate $\gamma_r$ of transitions $|a\rangle \rightarrow
|b_\pm\rangle$.

	We can understand the origin of this incoherent pumping
by considering a two-level system coupled to a radiation
reservoir.  The reduced density matrix operator derived in
Weisskopf-Wigner approximation~\cite{scullybook} has the form
\begin{eqnarray} \nonumber
\dot \rho (t) &=& -\bar n_{th} \gamma_r \left [ \hat \sigma_- 
\hat \sigma_+ \rho (t)
- \hat \sigma_+ \rho(t) \hat \sigma_- \right ] - \\
\label{densm}
&&
(\bar n_{th} + 1)\gamma_r\left [ \hat \sigma_+ \hat \sigma_- \rho (t)
- \hat \sigma_- \rho(t) \hat \sigma_+ \right ] +
\mathrm{H.c.}\,, 
\end{eqnarray}
where $\bar n_{th}$ is thermal average photon number in the
reservoir, $\gamma_r$ is the atomic decay rate of the upper
level, $\hat \sigma_- = |b \rangle \langle a|$ and $\hat
\sigma_+ = |a \rangle \langle b|$.  To model atomic excitation
by the incoherent radiation in the reservoir, the incoherent
pumping rate can be written as $R = 2\gamma_r \bar n_{th}$\,.
From Eq.~(\ref{densm}) we find the equations of motion for
the excited state population:
\begin{eqnarray} \label{rhoaa2}
\dot \rho_{aa} = -2 \gamma_r (\bar n_{th}+1) \rho_{aa} + 2 \gamma_r \bar n_{th}
\rho_{bb}~.
\end{eqnarray}
In an optically thin atomic medium the probability of photon
reabsorption is small and $\bar n_{th} = 0$.  However, in
optically thick media photons diffuse slowly and $\bar n_{th}
\ne 0$.  The value of $\bar n_{th}$ can be estimated from a
rate equation
\begin{equation} \label{rate}
\dot{\bar n}_{th} = - r_e \bar n_{th} + r_a \rho_{aa}~,
\end{equation}
where $r_e$ is the photon escape rate and $r_a$ is the pumping
rate due to the atomic decay.  Both $r_e$ and $r_a$ depend on
the geometry of the system and on the atomic density.  In steady
state, $\bar n_{th} = r_a \rho_{aa}/r_e$\,, and it follows from
Eq.~(\ref{rhoaa2}) and $\rho_{aa}+\rho_{bb}=1$ that $r_e > r_a$.

	It is convenient to formally introduce a function
$f(N)$ defined by $r_a/r_e = f/(1+f)$ which characterizes the
radiation trapping, such that $f(N)\geq 0$ and $f(N=0)=0$.
In the case when most population is in the ground state,
$\rho_{aa} \ll 1$, we see from Eq.~(\ref{rhoaa2}) that $\dot
\rho_{aa} \approx -2\gamma_r \rho_{aa}/(1+f)$.  In the limit of
low light intensity $|\Omega| \ll \gamma_r$ and weak radiation
trapping $\bar n_{th} \ll 1$ the cw light propagation obeys $d
|\Omega|^2/dz \simeq -2\kappa\gamma_r\rho_{aa}$, where $\Omega$
is the Rabi frequency of the transition, $\kappa=(3/8\pi)
N\lambda^2 \gamma_r$\,, $\lambda$ is the wavelength, and $z$
is the distance of propagation through the medium.  We thus
have a simple equation for the incoherent pumping rate due to
radiation trapping:
\begin{equation}
\label{r}
R = -\frac{1}{\kappa}\, \frac{f(N)}{1+f(N)} \,
\frac{{\rm d}}{{\rm d}z} |\Omega|^2\,.
\end{equation}
This is a very intuitively appealing model:  radiation trapping
can exist only if the coherent radiation is absorbed by the
system and is scattered due to spontaneous emission.

	We now return to the problem of radiation trapping in a
three-level system with EIT.  Because the Doppler distribution
depends on atomic density (temperature), it is simplest to study
radiation trapping in the Doppler-free limit of EIT, i.e., when
the absorption and the dispersion do not depend on the width
of the Doppler distribution $W_d$\,.  Doppler averaging shows
that this condition is fulfilled for relatively large light
intensities $|\Omega(z)| \gg W_d \sqrt{\gamma_0 /\gamma_r}$
for any $z$.

	The stationary propagation of the right and left
circular polarized electric field components through
the atomic vapor is described by Maxwell-Bloch equations
in the slowly-varying amplitude and phase approximation
\cite{scullybook}.  We solve the equations by considering only
the lowest order in $\gamma_0$, $R$ and $\delta_0$, assuming
$|\Omega_-(z)|^2 \approx |\Omega_+(z)|^2$, where $\Omega_\pm$
are the complex Rabi-frequencies of the two optical fields.
We separately consider the spatial evolution of the amplitudes
and phases of these complex Rabi-frequencies by writing
$\Omega_\pm(z) = |\Omega_\pm(z)| \mathrm{e}^{i\phi_\pm(z)}$,
and derive equations for the total intensity $|\Omega|^2$
and the relative phase $\phi=\phi_--\phi_+$
\begin{eqnarray}
\frac{{\rm d}}{{\rm d}z} |\Omega|^2 &=& -\kappa (\gamma_0 + R)\,,
\label{intensity1} \\
\frac{{\rm d}}{{\rm d}z} \phi &=& \delta_0 \frac{\kappa}{|\Omega|^2}\,.
\label{ph1}
\end{eqnarray}
To solve these equations we must specify the functional
form of the incoherent pumping rate $R$ in the three level
configuration.  From the general properties of radiation
trapping~\cite{Molish} and from the results from radiation
trapping in a two-level system, we assume that in the case
of Doppler-free EIT the incoherent pumping can be modeled by
Eq.~(\ref{r}) as in the case of a two-level system.  As we
shall see, this model works very well.

	With this form for $R$, Eq.~(\ref{intensity1}) can be
easily solved and we arrive at
\begin{eqnarray}
\left |\frac {\Omega(z)}{\Omega(0)} \right |^2 &=&
1-\frac{\gamma_0\kappa z}{|\Omega(0)|^2} \left(1+f(N)\right)
\label{intensity_sol}
\end{eqnarray}
so from Eq.~(\ref{r}) we have $R= f(N)\gamma_0$.  Integration of
Eq.~(\ref{ph1}) for the phase yields
\begin{equation}
\label{phi2}
\left. \frac{d \phi (z)}{dB}\right|_{B\rightarrow 0}
= \frac{2\mu_B}{\hbar\left(\gamma_0 + R\right)}
{\rm ln}\,\left |   
{\Omega(0) \over \Omega (z)}
\right |^2 .
\end{equation}
Detection of $|\Omega(L)/\Omega(0)|^2$ and $d\phi (L)/dB$,
allows us to infer the value of the coherence decay rate as
a function of the atomic density and estimate the radiation
trapping effect.  Thus we see that for optically thick media
the coherence decay rate increases with the density.

	For smaller intensities $|\Omega(z)| \ll W_d
\sqrt{\gamma_0 /\gamma_r}$\,, Doppler-free EIT is not
established, so the approximation Eq.~(\ref{r}) is not valid
and we do not discuss this regime here.

	Based on the low-density data in Fig.~\ref{fig2.fig}
(for which radiation trapping is negligible)
Eqs.(\ref{intensity_sol}) and (\ref{phi2}) allow us to
determine the coherence decay rate to be $\gamma_0 \approx
0.004 \gamma_r$\,.  Given this value, we can then use the
high density set of these data and Eq.~(\ref{phi2}) to obtain
the incoherent pumping rate $R$ due to radiation trapping.
The dependence is shown by the dots in Fig.~\ref{fig3.fig}.
In general, the functional form of $R$ is not an ``absolute''
and it changes if the cell geometry or laser beam size changes,
which is a key signature of the effect of radiation trapping.

	The probability of photon reabsorption becomes
significant when the medium becomes optically thick on the
length scale of the atomic cell size~\cite{Molish} (under the
Doppler-free EIT condition almost all atomic population is in
the ground state), or
\begin{equation} \label{un}
{3 \over 8\pi} N\lambda^2 d {\gamma_r \over W_d} > 1\,.
\end{equation} 
For our experiment $\gamma_r / \Delta_D \approx
0.01$, so Eq.~(\ref{un}) is fulfilled for $N>5 \times
10^{10}~\mathrm{cm}^{-3}$.  For densities less than this,
radiation trapping is negligible and we have $R \approx 0$.
Above this value there are two distinct regimes of behavior,
with both seen in Fig.~\ref{fig3.fig}.  At low density we
have that $R$ increases linearly with density due to photon
absorption and emission within the cell.  If the atomic beam
is narrower than the radius of the atomic cell, as we have
in our experiment ($d \sim 0.1D$), the next regime occurs
for densities when photon reabsorption becomes significant
inside the laser beam.  In our case this is $N>5 \times
10^{11}~\mathrm{cm}^{-3}$.

	To confirm our simple analytical calculations we have
also made detailed numerical simulations of the experiment.
We have considered light propagation in a thirteen-level
Doppler-broadened system corresponding to the $F=2 \rightarrow
F'=1,2$ transition in ${}^{87}$Rb.  The decay of the atomic
coherence was modeled by finite time of the flight through
the laser beam (an open system).  We solved the density
matrix equations in steady state using the coherence decay
rate as a fit parameter.  In other words, we choose the
effective coherence decay rate $\gamma_0 + R$ in such a way
that our numerical points for the dispersion $d\phi /dB$ and
intensity $I_{\mathrm{out}}/I_{\mathrm{in}}$ corresponds to
the experimental results.  This is shown in the solid line in
Fig.~\ref{fig2.fig}.  The dependence for $R/\gamma_0$ obtained
this way is shown in the solid line in Fig.~\ref{fig3.fig}.
We see that the simple analytical analysis of the data coincides
with the simulations for low atomic densities and diverges
slightly for high densities.  We explain this difference by
inadequate intensity of the laser light.  The maximum intensity
of our laser ($\sim 100~\mathrm{mW/cm}^2$) corresponds to a
Rabi frequency $|\Omega_0| \sim 3.6 \gamma_r$\,, which lies
on the edge of Doppler-free region determined by $|\Omega_0|
\geq W_d\sqrt{\gamma_0/\gamma_r} \approx 6 \gamma_r$\,.
The absorption further decreases the intensity resulting
in the Doppler broadening becoming important, unlike in our
simplified calculations.

	Finally, we note that the observations
reported here cannot be explained by spin exchange
collisions between the atoms.  The collisional
cross section for Rb atoms is approximately $2\times
10^{-14}~\mathrm{cm}^2$~\cite{happer72rmp} which results in a
coherence decay rate $\gamma_0 \approx 2\times 10^{-5}\gamma_r$
for the densities reported here.  This is approximately
two orders of magnitude less than the time-of-flight limited
coherence decay rate $\gamma_0 \approx 4\times 10^{-3}\gamma_r$
that we measured.

	In conclusion, we have shown both experimentally and
theoretically that the effect of radiation trapping enhances
the decay rate of the atomic coherence established by linearly
polarized laser radiation between Zeeman sublevels.  This effect
leads to significant increase of the residual absorption in
EIT experiments with optically thick atomic vapors.

	The authors  gratefully acknowledge useful discussions
with  D.\ Budker, M.\ Fleischhauer, L.\ Holberg, E.\ E.\
Mikhailov, Y.\ V.\ Rostovtsev, V. \ A. \ Sautenkov, V.\ L.\
Velichansky, R.\ Wynands, and V.\ Yashchuk, and the support from
the Office of Naval Research, the National Science Foundation,
and the Welch Foundation.

\def\ibid{{\em ibid.}}
\def\etal{{\em et al.}}

\newpage

\begin{figure}[ht]
 \vspace*{2ex}
 \caption{
 	\label{levels.fig}
	Diagram showing the experimental setup; inset:
	Idealized three-level $\Lambda$-scheme considered in
	the theoretical calculations.
}
\end{figure}

\begin{figure}[ht]
 \vspace*{2ex}
 \caption{
 	\label{fig2.fig}
	The dependence of rotation rate $d
	\phi/dB$ on transmission through the system
	$I_{\mathrm{out}}/I_{\mathrm{in}}$\,: experimental
	(dots), previous theory with $\gamma_0=0.004 \gamma_r$
	(dashed line), and obtained by numerical simulation
	including radiation trapping.  (solid line).
}
\end{figure}

\begin{figure}[ht]
 \vspace*{2ex}
 \caption{
 	\label{fig3.fig}
	The incoherent pumping rate $R/\gamma_0$ due to
	radiation trapping as a function of atomic density $N$:
	calculated by applying Eq.~(\ref{phi2}) to the data (dots)
	and obtained by numerical simulation (solid line).
}
\end{figure}

\newpage

\centerline{\epsfig{file=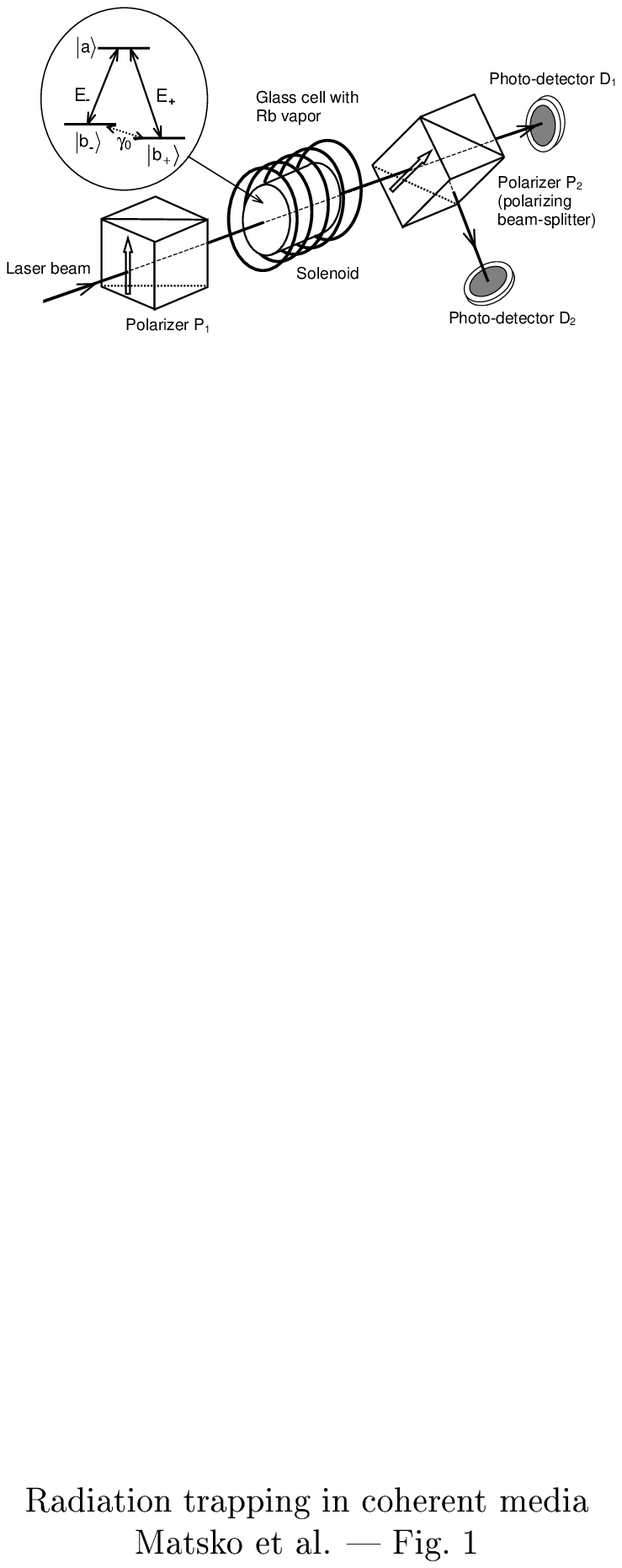}}
\centerline{\epsfig{file=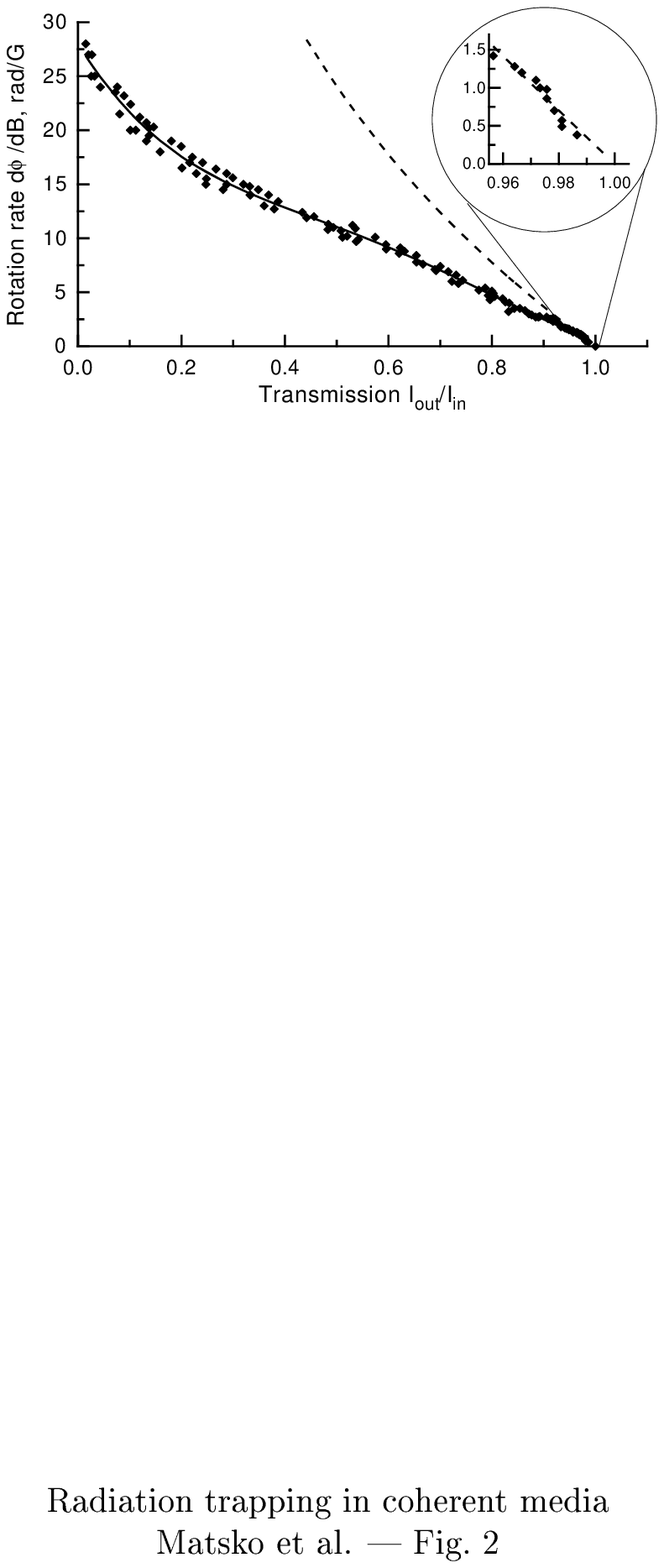}}
\centerline{\epsfig{file=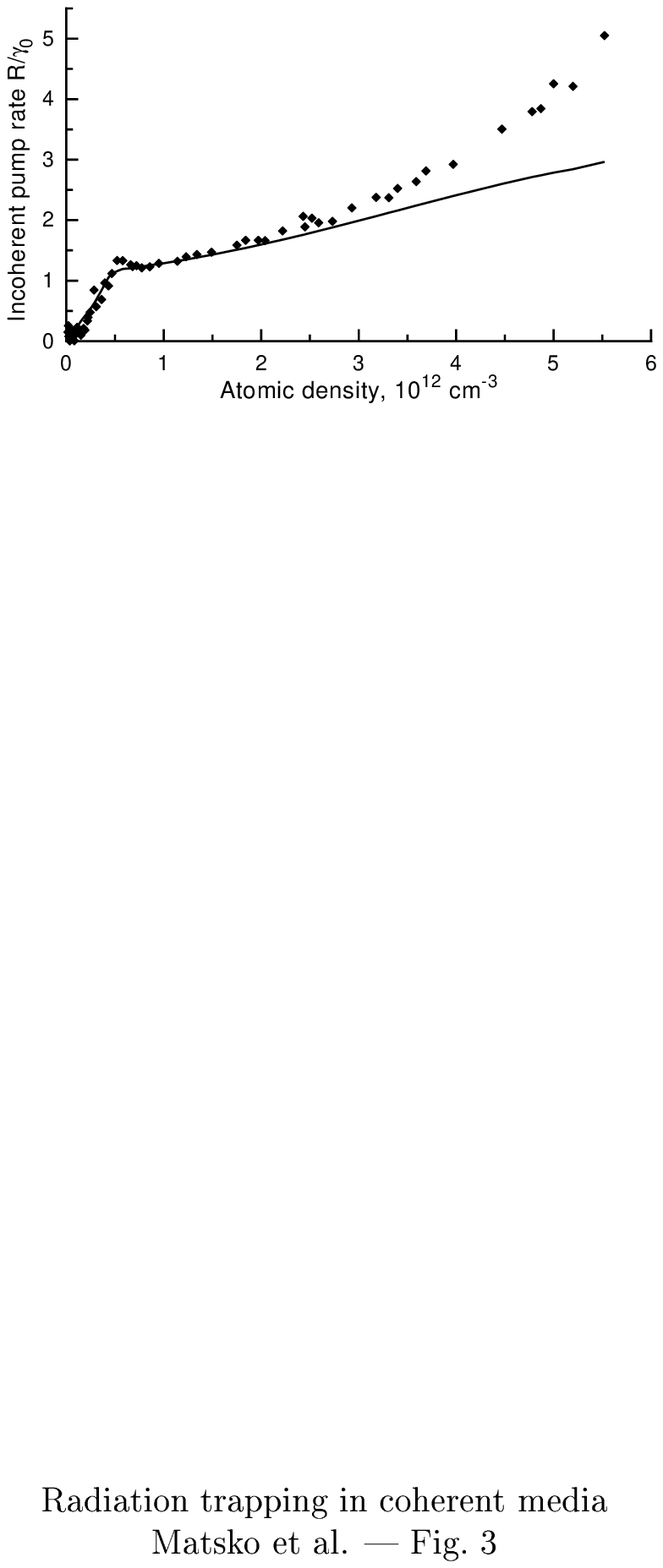}}

\end{document}